%% file: draft.tex
\documentclass[aps,11pt]{revtex4}
\usepackage{graphicx}
\usepackage{latexsym}
\usepackage{amsmath}
\usepackage{amsfonts}

\newcommand{\la}{\left<}
\newcommand{\ra}{\right>}
\newcommand{\veff}{\mbox{$v_\text{eff}$}}
\newcommand{\Nav}{\mbox{$\la N \ra$}}
\newcommand{\Rgid}{\mbox{$R^{(0)}_\text{g}$}}
\newcommand{\Rgyr}{\mbox{$R_\text{g}$}}
\newcommand{\Rend}{\mbox{$R_\text{e}$}}
\newcommand{\astar}{\mbox{$a^*$}}
\newcommand{\bstar}{\mbox{$b^*$}}
\newcommand{\rhostar}{\mbox{$\rho^*$}}

\begin{document}
\title{Intramolecular Form Factor in Dense Polymer Systems:\\
Systematic Deviations from the Debye formula}

\author{P. Beckrich$^1$, A. Johner$^1$, A. N. Semenov$^1$, 
S. P. Obukhov$^{1,2}$, H. Beno\^\i t$^1$ and J. P. Wittmer$^1$}
\affiliation{$^1$ Institut Charles Sadron,6 Rue Boussingault, 67083 Strasbourg Cedex, France}
\affiliation{$^2$ Department of Physics, University of Florida, USA}

\date{\today}

\begin{abstract}
We discuss theoretically and numerically the intramolecular form factor $F(q)$ 
in dense polymer systems. Following Flory's ideality hypothesis, chains in the melt 
adopt Gaussian configurations and their form factor is supposed to be given 
by Debye's formula.  
At striking variance to this, we obtain noticeable (up to 20\%) non-monotonic 
deviations which can be traced back to the incompressibility of dense polymer 
solutions beyond a local scale. 
The Kratky plot ($q^2F(q)$ {\it vs.} wavevector $q$) does not exhibit 
the plateau expected for Gaussian chains in the intermediate $q$-range.
One rather finds a significant decrease according to the correction
$\delta(F^{-1}(q)) = q^3/32\rho$ that only depends on the concentration 
$\rho$ of the solution, but neither on the persistence length or the
interaction strength. 
The non-analyticity of the above $q^3$ correction is linked to the existence 
of long-range correlations for collective density  fluctuations that survive
screening. 
%
Finite-chain size effects are found to decay with chain length $N$ as $1/\sqrt{N}$.
\end{abstract}
\maketitle

\centerline{PACS numbers: 05.40.Fb, 05.10.Ln, 61.25.Hq, 67.70.+n}

\newpage
\section{Introduction}
\label{sec_intro}

Following Flory's ideality hypothesis \cite{Flory}, one expects a macromolecule 
of size $N$, in a melt (disordered polymeric dense phase) to follow Gaussian
statistics \cite{scalingPGG,doi1989,khokhlov+grosberg,Rubinstein}. The official 
justification of this mean-field result is that density fluctuations 
are small, hence negligible.

Early Small Angle Neutron Scattering experiments \cite{BenBook,RawisoLectures} 
have been set up to check this central conjecture of polymer physics. 
The standard technique measures the scattering function $S(q)$ 
($q$ being the wave-vector) of a mixture of deuterated (fraction $f$) and 
hydrogenated (fraction $1-f$) otherwise identical polymers. 
The results are rationalized \cite{Boue,Rubinstein} {\it via} the formula
\begin{equation}
S(q)\propto f(1-f) F(q)
\label{labeledscatt}
\end{equation}
to extract the form factor (single chain scattering function) $F(q)$.
To reveal the asymptotic behavior of the form factor for a
Gaussian chain $F^{(0)}(q)\sim 12/q^2b^2$, one usually plots
$q^2 F({q})$ versus $q$ (called ``Kratky plot").
The aim would be to show the existence of the ``Kratky plateau" in the
intermediate range of wave-vectors 
$\frac{ 2\pi }{\Rgyr} \lesssim q \lesssim \frac{2\pi}{b}$ 
(``Kratky regime") 
where $\Rgyr$ is the radius of gyration of the macromolecule and 
$b$ is the (effective) statistical segment length \cite{doi1989}.
In contrast to the low-$q$ ``Guinier regime" ($q \lesssim \frac{ 2\pi }{\Rgyr}$),
clean scattering measurements can be performed in the Kratky regime 
\cite{footInhomogeneities}
suggesting the measurement of $b$ from the height of the Kratky plateau.

Surprisingly, this plateau appears to be experimentally elusive as already 
pointed out by Beno\^\i t \cite{BenBook}: ``Clearly, Kratky plots have to be
interpreted with care``. For typical experiments, the available
$q$-range is $5 \cdot 10^{-3}/\text{\AA} < q < 0.6/\text{\AA}$. 
Kratky plots are quickly increasing at high $q$
\cite{RawisoLectures,BenBook}, because of the rod-like effect starting
at $q\sim \frac{1}{l_\text{p}}$, when
the beam is scanning scales comparable with the persistence length
$l_\text{p}$ (this regime is in fact used
to assess $l_\text{p}$). Sometimes these curves can also quickly
decrease, this is usually attributed to the fact
that the chain cannot be considered as infinitely thin. The finite
cross section of the
chain tends to switch off the signal
\cite{RawisoLectures,BenBook} as
$\exp\left({-\frac{q^2 R^2_\text{c}}{2}}\right)$, where $R_\text{c}$
is the radius of gyration of the cross section (in CS$_2$, a
good solvent for dilute polystyrene, $R^2_\text{c}=9.5\text{\AA}^2$
\cite{MRRDCP1987}).
Rawiso {\it et al.} \cite{RawisoLectures,BenBook}
have also shown that sometimes these two effects can compensate, for
instance in a blend
of hydrogenated and fully deuterated high molecular weight
polystyrene, letting appear a {\it pseudo} Kratky plateau, 
extending outside the intermediate regime to higher $q$-values. 
In fact, taking for instance the case of polystyrene with 
$l_\text{p}\simeq 10\text{\AA}$, all these parasitic effects never 
really allowed neutron scattering experiments to confirm the Flory's hypothesis.
Up to now {\em no} scattering experiment has been performed on a sample 
allowing for a test over a wide enough range of $q$ \cite{footqrange,Bates}
and  there exists {\em no} clear experimental evidence of 
the Kratky plateau expected for Gaussian chains.

As we will show in this paper, there are fundamental reasons why this plateau may actually 
{\em never} be observed, even for samples containing very long and flexible polymers.
Recently, long-range correlations, induced by fluctuations, have been theoretically derived 
\cite{ANSAJ2003,jojoPRL,ANSSO2005,SOANSPRL2005,papEPL,BeckrichThesis} and numerically 
tested for two \cite{cavallowittmer} and three-dimensional \cite{jojoPRL,papEPL} 
dense polymer systems.
(Similar deviations from Flory's ideality hypothesis have been also reported in 
various recent numerical studies on polymer melts \cite{CSGK91,Auhl03} and networks 
\cite{Sommer05,SGE05}.)
The conceptually simpler part of these effects is related to the correlation hole 
\cite{scalingPGG} and happens to dominate the non-Gaussian deviations to the form 
factor described here \cite{footMFcycles}.

This is derived in the following Section~\ref{sec_analytic}.
There we first recapitulate the general perturbation approach (Sec.~\ref{sub_perturb}),
discuss then the intramolecular correlations in Flory size-distributed
polymers (Sec.~\ref{sub_intraFlory}). We obtain the form factor
of monodisperse polymer melts by inverse Laplace transformation of the
polydisperse case (Sec.~\ref{sub_intraMono}). 
In Section~\ref{sec_comput} these anal\-yti\-cal predictions are illustrated 
numerically by Monte Carlo simulation of the three dimensional bond-fluctuation model 
\cite{BFM,BWM04}. We compare melts containing only monodisperse chains
\cite{jojoPRL,papEPL}
with systems of (linear) equilibrium polymers (EP) \cite{CC90,WMC98,HXCWR06}
where the self-assembled chains (no closed loops being allowed)
have an annealed size-distribution of (essentially) Flory type
\cite{footquench2anneal}.
Excellent parameter free agreement between numerical data and theory is 
demonstrated, especially for long EP (Sec.~\ref{sub_simuEP}).
Finite-chain size effects are addressed as well.
In the final Section~\ref{sec_conclusion} the experimental situation is 
reconsidered in the light of our analytical and computational results.
There we show numerically that eq~\ref{labeledscatt} remains an accurate 
method for determining the form factor which should allow to detect 
the long-range correlations experimentally.
%

\newpage
\clearpage
\section{Analytical Results}
\label{sec_analytic}

\subsection{The Mean-Field Approach}
\label{sub_perturb}

It is well accepted that at the mean-field level \cite{doi1989}, 
the excluded volume interaction is entropically screened in dense 
polymeric melts.  Long ago Edwards and de Gennes 
\cite{doi1989,scalingPGG,RPAEd1,RPAEd2,cloiz}
developed a self-consistent mean-field method to 
derive a screened mean (molecular) field: 
this theory is an adaptation of the Random Phase Approximation (RPA) \cite{Nozieres} to 
polymeric melts and solutions. The famous result of this approximation gives the response
function $S(q)$ as a function of $F^{(0)}(q)$, the scattering
function of a Gaussian (phantom) chain {\it via} the relation:
\begin{equation}
\frac{1}{S(q)} = \frac{1}{\rho F^{(0)}(q)} + v,
\label{eq_RPA}
\end{equation}
where $\rho$ is the mean concentration of monomers in the system and
$v$ is the bare excluded volume 
(proportional to the inverse of the compressibility of the system). 
Please note that $F^{(0)}(q)$ 
has to be properly averaged over the relevant size-distribution 
of the chains \cite{cloiz}. 
To get the effective interaction potential between monomers, we label
a few chains. The interactions between labeled monomers are screened
by the background of unlabeled monomers. Linear response gives
the effective $q$ dependent excluded volume:
\begin{equation} \label{effvol}
\frac{1}{\veff(q)}= \frac{1}{v} + \rho  F^{(0)}(q).
\end{equation}

Let us from now on consider a dense system of long chains with exponentially 
decaying number density $\rho_N = \rho \mu^2 e^{-\mu N}$
for polymer chains of length $N$ with $\mu \equiv 1/\Nav$ 
being the chemical potential.
This so-called Flory distribution is relevant to EP systems \cite{CC90,WMC98}.  
Hence, eq~\ref{effvol} yields (using eqs~\ref{defpoly},\ref{eq_idealform} indicated below)
\begin{equation} \label{potential}
\veff(q) \hspace{0.1cm}
                  \underset{R_\text{g}  \gg \xi  }{\approx}
\hspace{0.1cm}  \frac{v \xi^2 }{1 + q^2 \xi^2 }(q^2 +
\frac{\mu}{a^2}).
\end{equation}
Here $a$ is the characteristic length of the monomer ($a^2 \equiv {b^2}/{6}$) and 
$\xi=\sqrt{\frac{a^2}{2 \rho v}}$ is the mean-field correlation length.
When chains are infinitely long ($\mu\rightarrow 0$) we recover the
classical result by Edwards \cite{doi1989} ignoring finite-size
effects. If we further restrict ourselves to length scales larger
than $\xi$ ($q\xi\ll 1$) eq~\ref{potential} simplifies to
$\veff(q)\approx q^2a^2/2\rho$ which does not depend on the bare
excluded volume $v$ and corresponds to the incompressible melt limit. 
For very large scales ($q \Rgyr \ll 1$) one obtains the contact
interaction associated to the volume $v^* \equiv \veff(q\rightarrow 0)$, 
such that $v^*/v =\mu \xi^2/a^2 \sim \xi^2/R_\text{g}^2$ 
(far weaker than the initial one given
in the direct space by $v({\bf r})= v \delta({\bf r})$). The
interaction $v^*$ is relevant to the swelling of a long chain
immersed in the polydisperse bath \cite{scalingPGG}.

The screened excluded volume interaction eq~\ref{potential} taken at
scale $\Rgyr$ is weak and decays with chain length  as $1/\Nav$.
The associated perturbation parameter $u$ in $d$-dimensional space depends
on chain length as $u \sim v^* N^2/ R_\text{g}^d \sim \Nav^{1-d/2}$ and 
the screened excluded volume potential is, hence,
perturbative in three dimensions \cite{scalingPGG}.

\begin{figure}[t]
\includegraphics*[width=10cm]{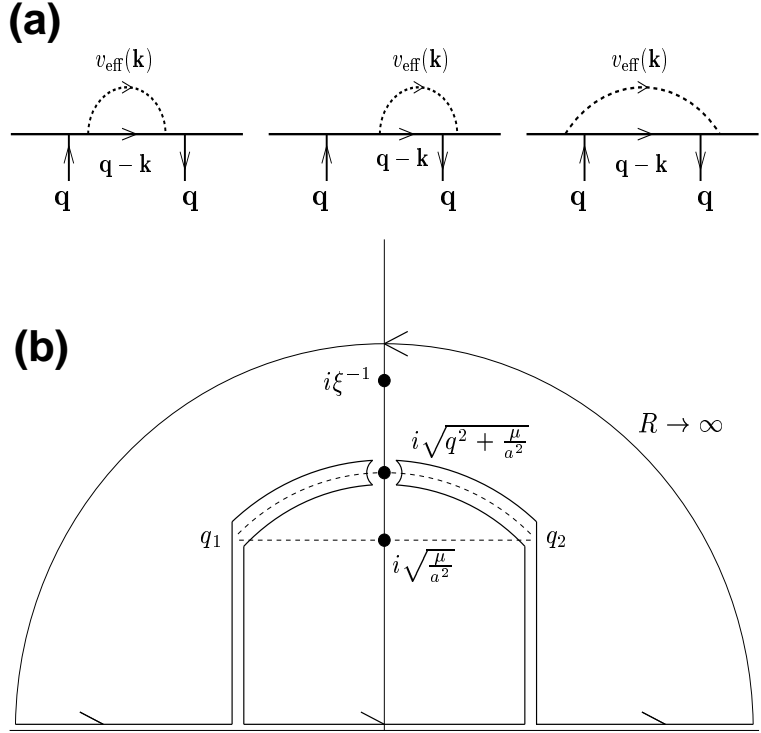}
\caption{{\bf (a)} Interaction diagrams and {\bf (b)} integration contour 
and singularities.
{\bf (a)}
There exist three non-zero contributions to first-order perturbation,
the first involving two points inside the segment of size
$s=\arrowvert i-j \arrowvert$ monomers, the second involving one
point inside and one outside the segment and the third involving one
point on either side of the segment. 
Momentum $q$ flows from one correlated point to the other. Integrals
are preformed over the momentum $k$. Dotted lines denote the
effective interactions $v_{eff}(k)$, bold lines the propagators.
{\bf (b)}
A possible contour in the complex plane. The dashed line depicts the
logarithmic branch-cut, which is an arc of the circle of radius
$r =\sqrt{\mu + q^2 a^2}$ between $q_1 = r e^{i(\pi-\alpha)}$
and $q_2 = r e^{i\alpha}$, with $\alpha = \arctan(\sqrt{\mu/q^2a^2})$.
The other singularities are single poles at $k=i\xi^{-1}$,
$k=i\sqrt{\mu/a^2}$ and at $k=i\sqrt{q^2 + \mu/a^2}$.
}
\label{fig1}
\end{figure}

Let us define $G({\bf q}, s) = \langle  \exp(-i {\bf q \cdot r}_{s})
\rangle$, the Fourier transform of the two-point intramolecular
correlation function, with $s=\arrowvert i-j \arrowvert$, the number
of monomers between the two positions
separated by ${\bf r}_{s}$.
One can perturb the two-point Gaussian correlation function
\begin{equation}
G^{(0)}({\bf q}, s) = \langle  \exp(-i {\bf q \cdot r}_{s}) \rangle_0 =
\exp{(-s  q^2  a^2)}
\label{eq_TwoPointGauss}
\end{equation}
with the molecular field eq~\ref{potential}. 
In this type of calculations, there are only three non-zero contributions \cite{ON83,Duplantier86}.
They are illustrated by the diagrams given in Figure~\ref{fig1}(a). 
Knowing this correlation, it is possible to derive many single-molecule properties
\cite{BeckrichThesis}.

\subsection{Intramolecular correlations for Flory size-distributed polymers\label{sub_intraFlory}}

\begin{figure}[t]
\includegraphics*[width=11cm]{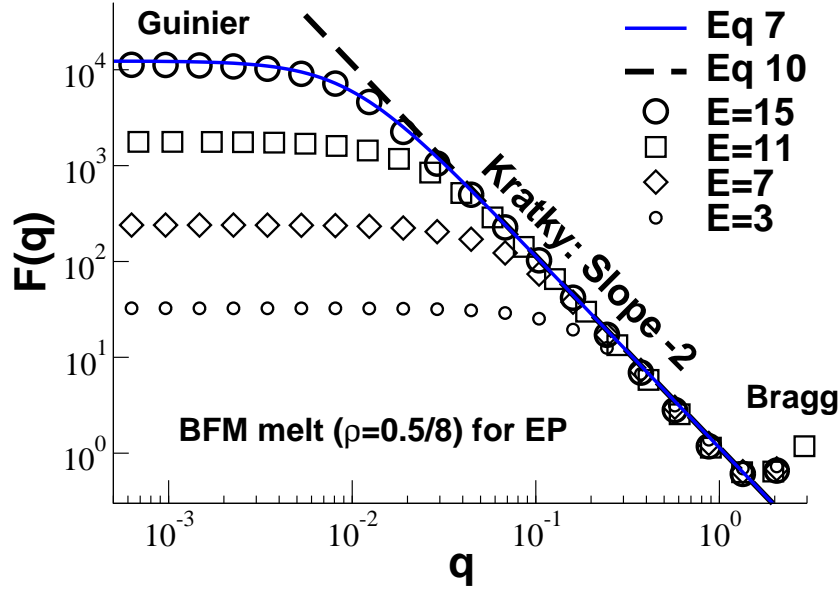}
\caption{
The intramolecular form factor $F(q)$ of polymer chains is an important
property since it allows to make a connection between theory and simulation
on the one hand and experiments of real systems on the other hand.
The ideal chain form factor for Flory size-distributed polymers,
eq~\ref{eq_idealform}, is indicated by the solid line.
In the Kratky regime between the total chain and monomer sizes the form factor
expresses the fractal dimension of the Gaussian coil, eq~\ref{eq_kratkyregime}
(dashed line). Experimentally, this is the most important regime since it
is, for instance, not affected by the (a priori unknown) polydispersity.
The form factors of equilibrium polymers (EP) of various scission energies $E$
obtained numerically are indicated.
The form factor of polydisperse polymer systems is obtained by computing
for each chain
$(\sum_i\sin({\bf q \cdot r}_i))^2 + (\sum_i\cos({\bf q \cdot r}_i))^2$,
summing over all chains (irrespective of their length) and dividing
by the total number of particles. The computational data reveal an
additional regime at wavevectors corresponding to the monomer structure
(``Bragg regime") which is not treated by our theory.
All data have been obtained for a number density $\rho=0.5/8$ of the three
dimensional bond-fluctuation model (BFM).
}
\label{fig2}
\end{figure}

\begin{figure}[t]
\includegraphics*[width=11cm]{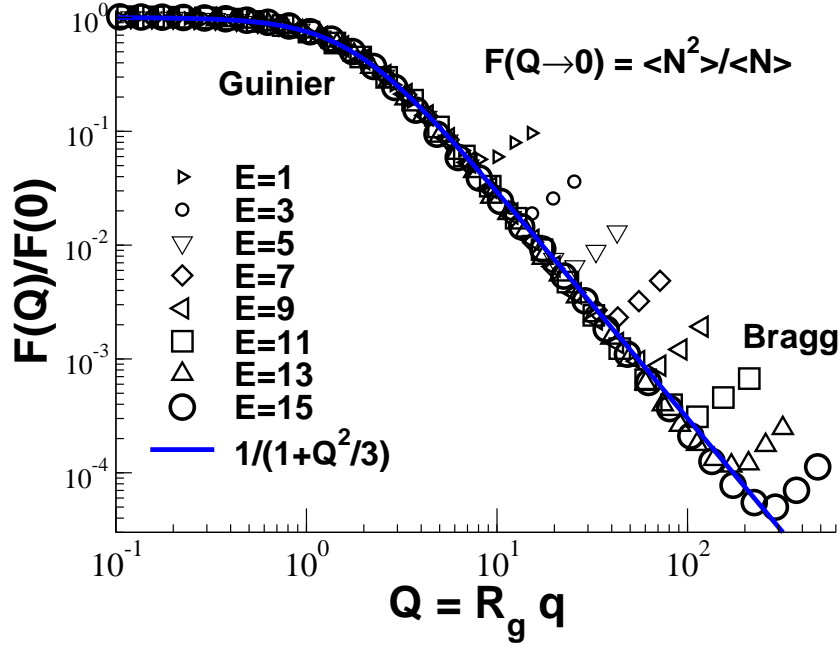}
\caption{Successful scaling of the form factors of EP obtained from
our BFM simulations for various scission energies $E$ as indicated:
$F(Q)/F(0)$ is plotted {\em vs.} the reduced wavevector $Q = \Rgyr q$.
Note that both scales $F(Q\rightarrow 0) = \la N^2 \ra/ \Nav$
and the (Z-averaged) gyration radius $\Rgyr$
have been directly measured for each sample
and are indicated in Table~\ref{tab_EP}.
Obviously, the scaling breaks down due to local physics for large wavevectors
(Bragg regime).
The line represents the prediction for ideal Flory-distributed polymers,
eq~\ref{eq_idealform}, with parameters chosen in agreement with the
Guinier limit, eq~\ref{eq_guinier}.
Importantly, in the intermediate Kratky regime small, albeit systematic,
deviations are visible which will be further investigated below.
}
\label{fig3}
\end{figure}

\begin{figure}[t]
\includegraphics*[width=11cm]{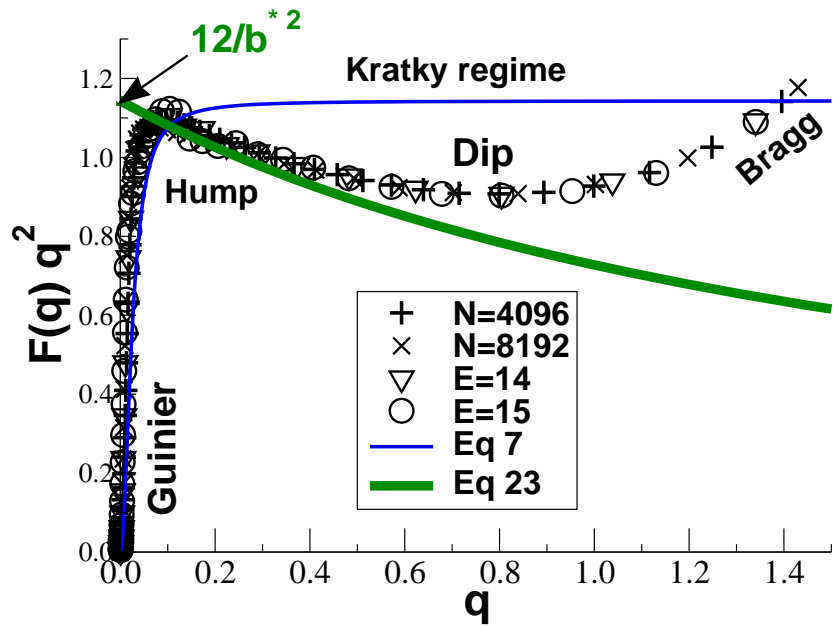}
\caption{
Kratky representation of the intramolecular form factor $F(q) q^2$ {\em vs.}
wavevector $q$ for monodisperse (crosses) and equilibrium polymers.
The {\em non-monotonous} behavior predicted by the theory is clearly demonstrated.
The ideal chain form factor, eq~\ref{eq_idealform} (thin line), overpredicts
the dip of the form factor at $q\approx 0.7$ by about $20\%$.
The bold line indicates the prediction for infinite chains,
eq~\ref{OneLoopForm},
which should hold for both system classes for infinitely long chains.
For this reason we have chosen the largest chains currently available.
%
}
\label{fig4}
\end{figure}

The intramolecular correlation function is investigated through its Fourier transform, 
the form factor. As already mentioned, we consider Flory size-distributed polymer systems
\cite{footquench2anneal}. In this case, one can define the form factor as:
\begin{equation}\label{defpoly}
F({q}) =   \sum_{N=0}^{\infty} \mu^2 \exp(-\mu N) \sum_{i=1}^N
\sum_{j=1}^N \langle
                  \exp(-i {\bf q \cdot r}_{i,j}) \rangle
\end{equation}
with ${\bf q}$ the scattering vector. If the chains followed
Gaussian statistics (as suggested by the Flory's hypothesis), 
one should find using eq~\ref{eq_TwoPointGauss}:
%
\begin{equation}
\label{eq_idealform}
F^{(0)}({q})= 
\frac{2}{\mu} \ \frac{1}{1 + q^2 a^2/\mu}.
\end{equation}
The ideal chain form factor for Flory size-distributed polymers is 
represented in the Figures~\ref{fig2}, \ref{fig3} and \ref{fig4} 
where it is compared to our computational results on EP discussed 
below in Section~\ref{sec_comput}.
In the small-$q$ regime, $q R_\text{g} \ll 1$ and for a polydisperse
system, we can measure $R_\text{g}$ {\it via} the Guinier
relation \cite{khokhlov+grosberg,cloiz,BenBook}:
\begin{equation}
F({q}) \hspace{0.05cm} \underset{q R^{(0)}_\text{g} \ll 1}\simeq 
\hspace{0.05cm}
\frac{\langle N^2 \rangle}{\Nav}
\left(1 - \frac{q^2 R_\text{g,Z}^{(0)\hspace{0.05cm} 2}}{3}\right).
\label{eq_guinier}
\end{equation}
Please note the Z-averaging \cite{cloiz} in the definition of $R_\text{g,Z}$ 
\begin{equation}
R_\text{g,Z}^2 = \frac{\sum_N N^2 \rho_N {R^2_{\text{g}}(N)}}{\sum_N N^2 \rho_N},
\end{equation}
$R_{\text{g}}(N)$ being the radius of gyration in the monodisperse case.
(Since only Z-averaged length scales are considered below 
the index Z is dropped from now on.) As the Flory distribution gives 
$\langle N^p \rangle = \Gamma(p+1)/\mu^p$, one has 
${R_\text{g}^{(0)\hspace{0.05cm} 2}}={3 a^2}/{\mu}$.
In the Kratky regime between coil and monomer size one
recovers the classical result for infinite chains
\begin{equation}
\label{eq_kratkyregime}
F^{(0)}({q}) \hspace{0.05cm} \underset{q R^{(0)}_\text{g} \gg 1}\simeq
\hspace{0.05cm} \frac{2}{q^2 a^2},
\end{equation}
(indicated by the dashed line in Figure~\ref{fig2}) 
which expresses the fractal dimension of the Gaussian coil.

Perturbing the Gaussian correlation function with the screened potential
$\veff({\bf r})=\int \veff(q) e^{i{\bf q \cdot r}}d^3{\bf q}/(2 \pi)^3$,
one has to evaluate
$\la \delta {\cal A} \ra = - \la U {\cal A} \ra_0 + \la U \ra_0 \la {\cal A} \ra_0$
with $U = \frac{1}{2}\sum_{i,j} \veff(r_i - r_j)$ for the observable ${\cal A}$ of interest
\cite{doi1989}. In our case, $\cal A$ corresponds the form factor, eq~\ref{defpoly} 
(without average). 
The calculation in reciprocal space is schematically illustrated by the three diagrams given 
in Fig.~\ref{fig1}(a) where bold lines represent propagators and dotted lines the effective 
interactions $\veff({\bf k})$. 
For the first perturbation contribution, $\la U F \ra_0$, the propagator 
carrying a momentum ${\bf p}$
(with ${\bf p} = {\bf q}$ or ${\bf p} = {\bf q} -{\bf k}$ as indicated in the diagrams)
corresponds to a factor $1/(p^2 a^2 + \mu)$. 
For the second contribution,
$\la U \ra_0 \la F \ra_0$, the same rules apply but ${\bf q}$ 
has to be replaced by zero and (then) $\mu$ by $\mu + q^2a^2$.
The momentum ${\bf k}$ is integrated out. Each diagram gives a converging contribution. 
Summing up the contributions of the diagrams (the central one has to be counted twice) and 
performing the angular integral (over the angle between ${\bf k}$ and ${\bf q}$) we arrive 
at the following integral:
\begin{equation} \label{total}
\begin{split}
\delta& F({q})=    F({q})-  F^{(0)}({q}) =\\
     & \frac{v}{4\pi^2 a^2(\mu + q^2 a^2)^2} \int_\mathbb{R} dk
\frac{k^2 \xi^2}{1+k^2 \xi^2} \hspace{0.1cm}
\frac{(2 \mu + (q^2+ k^2)a^2)^2 }{\mu+k^2 a^2} \\
                   &    \left(\frac{2}{\mu+(q^2+k^2) a^2}-\frac{1}{2 k q a^2}
\ln\left(\frac{\mu + (q + k)^2 a^2}{\mu + (q - k)^2 a^2}\right)\right).
\end{split}
\end{equation}
The contributions to this integral come from two poles, one at
$k=i\xi^{-1}$, this high-$k$ contribution renormalizes the
statistical segment, and one at $k=i \sqrt{q^2+\frac{\mu}{a^2}}$, and
from the logarithmic branch cut. The integration contour and the singularities
are illustrated in Figure~\ref{fig1}(b).
Absorbing the high-$k$ pole contributions in the renormalized
statistical segment $\bstar$ \cite{doi1989}, and using $\bstar$ instead of $b$ 
in the definition of $F^{(0)}({q})$, one finds (in the limit $q \xi < 1$)
as a function of $Q=q \Rgid = 3 q a^2/\mu$ 
\begin{equation} \label{totalpert}
\delta F(Q) / c = 
\frac{\sqrt{3}}{Q^2} \left(\frac{1}{(1+
\frac{Q^2}{3})^{3/2}} - 1\right)  
                    -\frac{1}{Q} \arctan \left(\frac{Q}{2
\sqrt{3}}\right) +\frac{1}{\sqrt{3}}
\frac{2 + Q^2}{(1+\frac{Q^2}{3})^2}.
\end{equation}
We have introduced here the factor $c = \frac{9 \Rgid}{\pi \rho \bstar^{4}}$
to write the deviation in a form which should scale with respect to chain length.
The statistical segment length $\bstar$, we have introduced, is given by
\begin{equation}\label{RenormStatSeg}
\bstar^2=b^2\hspace{0.05cm} \left(1+ \frac{12 v \xi}{\pi b^4} \right).
\end{equation}
which is consistent with the result obtained by Edwards and Muthukumar 
\cite{E75,ME82,doi1989}.

\begin{figure}[t]
\includegraphics*[width=11cm]{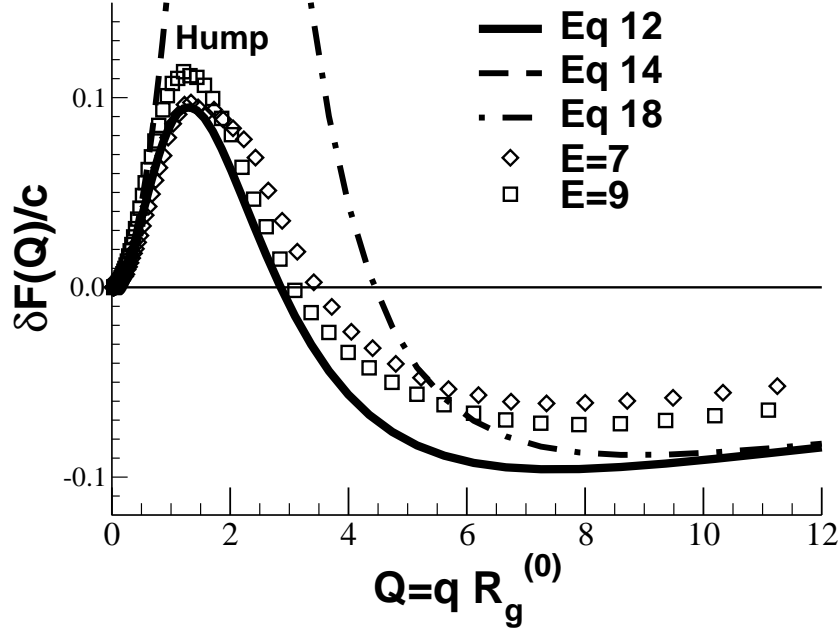}
\caption{Correction $\delta F(Q)=F(Q)-F^{(0)}(Q)$ to the form factor
as a function of $Q=q \Rgid$ (using $\Rgid^2 = 3 \astar^2/\mu$)
as predicted by eq~\ref{totalpert}
for Flory-distributed polymers. The deviation is positive for small
wavevectors (Guinier regime) and becomes negative at about $Q\approx 3$.
The scaling factor $c=9 \Rgid/\pi \rho \bstar^4$ should allow
a data collapse irrespective of the mean chain size --- provided
that the chains are sufficiently long to suppress additional physics
in the Bragg regime.
Also included are eq~\ref{guiniermu} and \ref{kratkymu} for the
asymptotic behavior in the Guinier and the Kratky regimes respectively.
The data from our BFM simulations (given here for two scission energies
where high precision data are available) agree quantitatively (especially
for small $Q$) with eq~\ref{totalpert}. (Note that the chains are too
short to allow a better fit for large $Q$.) The reference form factor $F^{(0)}(Q)$
has been computed from eq~\ref{eq_idealform} supposing a perfect Flory
distribution.
}
\label{fig5}
\end{figure}

As may be seen from the plot of eq~\ref{totalpert} in Figure~\ref{fig5}
the deviation from ideality is positive for small wave-vectors 
(with a pronounced maximum at $Q \approx 1$). It becomes negative
when the internal coil structure is probed ($Q \gg 2$).
Asymptotically, eq~\ref{totalpert} gives 
\begin{equation} \label{guiniermu}
\delta F(q) \underset{\frac{qa}{\sqrt{\mu}} \ll 1}\simeq
\frac{11 R^{(0)}_\text{g}}{4 \sqrt{3} \pi \rho \hspace{0.05cm}
\bstar^4} q^2 {R^{(0)\hspace{0.05cm} 2}_\text{g}},
\end{equation}
(dashed line in Figure~\ref{fig5})
which highlights the average swelling factor of the molecule in the melt:
\begin{equation}
\label{gyration}
R^2_\text{g} = {R^{(0) 2}_\text{g}}\hspace{0.05cm}\frac{\bstar^2}{b^2} \left( 1-
\frac{11 \sqrt{6}}{16 \pi \rho \hspace{0.05cm} \bstar^3}
{\sqrt{\mu}}  \right)
\end{equation}
(comparable to a logarithmic term in the two-dimensional case 
\cite{ANSAJ2003,NO81}). 
This is a sign of swelling, because  $R_\text{g} = \langle
N\rangle^{1/2} f(\sqrt{\langle N \rangle})$, with $f$, an increasing
function,
showing that the apparent swelling exponent $\nu$ for finite \Nav \
is slightly larger than $1/2$.
It is interesting to compare it with the (also Z-averaged) end-to-end 
distance, easily available because it involves only 
the top-left diagram of Figure~\ref{fig1}: 
\begin{equation}\label{endtoend}
R_\text{e}^2={R^{(0) 2}_\text{e}} \hspace{0.05cm}
\frac{\bstar^2}{b^2} \left( 1-\frac{10 \sqrt{6}}{16\hspace{0.05cm} \pi 
\rho \hspace{0.05cm} \bstar^3}
\sqrt{\mu} \right).
\end{equation}
(The diagram must be twice differentiated with respect to $\bf{q}$.)
The naively defined size dependent effective statistical segment of an $N$-chain 
(from $R_\text{e}^2=b^2_{\text{eff}}N$) therefore is:
\begin{equation}\label{EffStatSeg}
b^2_{\text{eff}} = \bstar^2\left(1 - \frac{\mbox{const}}{\rho \hspace{0.05cm} 
\bstar^3 \sqrt{N}} \right)
\end{equation}
The size-dependences in eqs~\ref{gyration},\ref{endtoend} follow
the same scaling, but the numerical factors are different.
Although internal segments carry a smaller correction \cite{jojoPRL}, 
the size-dependent contact potential  $\mu/2 \rho$ in eq.~\ref{potential}
counterbalances this effect and makes the correction to eq~\ref{endtoend} 
a little smaller than the one in eq~\ref{gyration}.

The asymptotic behavior of eq~\ref{totalpert} in the Kratky regime gives
\begin{equation} \label{kratkymu}
\delta F(q) \underset{q R_\text{g} \gg 1}\simeq \frac{12}{q^2 \bstar^2}\left(
\frac{3\sqrt{6}}{\pi \rho \hspace{0.05cm} \bstar^3} \sqrt{\mu} -
\frac{3 q}{8\rho \hspace{0.05cm} \bstar^2}\right),
\end{equation}
which is represented in Figure~\ref{fig5} by the dashed-dotted line.
The first term in this equation is a size-dependent shift of the Kratky plateau, 
and the second one, independent of the size, makes the essential difference with 
the Flory prediction (bold line in Figure~\ref{fig4}). Hence, the corrections 
induced by the screened potential are non-monotonic (Figure~\ref{fig5}). 

Eq~\ref{total} is not restricted to $q \xi < 1$,  it is applicable over the entire 
$q$-range in the case of weakly fluctuating dense polymers \cite{footcumbersome}
(mean-field excluded volume regime) as may be simulated with soft monomers 
allowed to overlap with some small penalty. 
In the case of strong excluded volume and
less dense solutions (critical semidilute regime, not explicitly
considered here) the results are valid at scales larger than $\xi$
provided the statistical segment is properly renormalized.
Quantitatively, the Ginzburg parameter measuring the importance
of density fluctuations reads $ G_z^2=v/(\rho\hspace{0.05cm} b^6)$.
For persistent chains $l_\text{p}>t$, $t$ being the thickness of the chain,
density fluctuations are negligible provided
$G_z^2\sim 1/(\tilde \rho\hspace{0.05cm} l_\text{p}^3/t^3) \ll 1$,
with $\tilde \rho = \rho\hspace{0.05cm} t^3$, the monomer volume 
fraction. The above makes
sense if $\xi \gg l_\text{p}$, which requires $\tilde \rho < t/l_p$.
This criterion also indicates the isotropic/nematic transition
\cite{khokhlov+grosberg,LiqCrysSem}. In summary, mean-field applies provided
$\left(\frac{t}{l_\text{p}}\right)^3 <\tilde \rho < \frac{t}{l_\text{p}}$.

\subsection{Monodisperse polymer melts in three dimensions}
\label{sub_intraMono}

It is possible to relate the form factor of the polydisperse system 
(Flory distribution) to the form factor $F_N({q})$  of a monodisperse system.  
Following eq~\ref{defpoly}, 
\begin{equation}
F(q) = \frac{\sum_{N=0}^{\infty} \rho_N N F_N({q})}{\sum_{N=0}^{\infty} \rho_N N}
=  \sum_{N=0}^{\infty} \mu^2 \exp(-\mu N) N F_N({q})
\label{eq_mono2poly}
\end{equation}
the deviations of the form factors of monodisperse and polydisperse
systems are related by the inverse Laplace transformation
$ \delta F_N({q})=\frac{1}{N}
\mathcal{L}^{-1}\left(\frac{1}{\mu^2} \delta F(q) \right)$,
$\mathcal{L}$ being the Laplace transform operator. 
Using our result eq~\ref{totalpert} for polydisperse systems this yields:
\begin{eqnarray} \label{eq_mono}
\delta F_N(Q)/c & \simeq  & 
          \frac{2}{\sqrt{\pi} Q^2}  e^{-Q^2}
           + \frac{5}{\sqrt{\pi} Q^2} 
           - \left( \frac{2}{\sqrt{\pi} Q} +  \frac{7}{\sqrt{\pi} Q^3} \right) D(Q) 
\nonumber
   \\ 
     &  + & \frac{4}{\sqrt{\pi} Q^3} \int_{0}^{1}\left( 
\frac{2}{\alpha^3}D\left(\frac{\alpha Q}{2} \right)
- \frac{Q}{\alpha^2}\right) d\alpha ,
\end{eqnarray}
where $Q = q R^{(0)}_{\text{g},N}$ with $R^{(0)}_{\text{g},N} = \sqrt{N} a$, 
the radius of gyration in the ideal monodisperse case and
$D(x)$ is the Dawson function, whose definition is 
$D(x)= e^{-x^2} \int_0^x e^{t^2} dt$ \cite{abramowitz}.
\begin{figure}[t]
\includegraphics*[width=11cm]{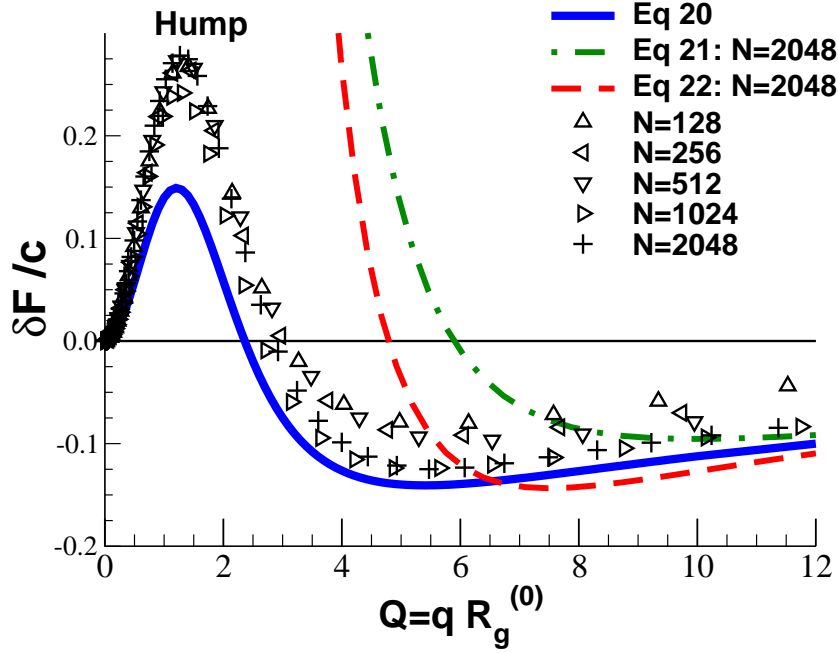}
\caption{Deviation to the ideal form factor in a monodisperse system,
eq~\ref{eq_mono}.
The shape is similar to that in the polydisperse case (Figure~\ref{fig5})
since polydispersity merely affects the Guinier regime.
Also indicated are the asymptotic behavior in the Kratky regime,
eq~\ref{eq_monokratky}, and the prediction for infinite chains,
eq~\ref{eq_infformfac}.
The simulation data scales roughly, especially in the Guinier
limit (hump), while for larger $q$ deviations are visible which may be
attibuted to local physics.
The agreement with eq~\ref{eq_mono} is only qualitative here since a much more
proncounced non-monotonic behavior is seen in the simulation. This is due to use
of the {\em polydisperse} chain perturbation potential, eq~\ref{potential} ---
an approximation which must become insufficient for low wavevectors where the
coil size matters.
}
\label{fig6}
\end{figure}
Eq~\ref{eq_mono}, shown in Figure~\ref{fig6} (bold line),
is accurate up to the finite size corrections to the interaction potential 
as these have been calculated for the Flory distribution, eq~\ref{potential}.
However, on small length scales, this influence is weakened, and
in this limit, it gives:
\begin{equation} \label{eq_monokratky}
F(q)\underset{q R_\text{g}\gg 1}\simeq \frac{12}{q^2 \bstar^2}\left(1+ 
\frac{6 \sqrt{6}}{\pi^{3/2}
\rho \hspace{0.05cm} \bstar^3}
\frac{1}{\sqrt{N}} - \frac{3 q}{8 \rho \bstar^2} \right).
\end{equation}
Eq~\ref{eq_monokratky} is represented in Figure~\ref{fig6} by the 
dashed-dotted line \cite{footPade}. 
Taking $\langle N \rangle=1/\mu$, eq~\ref{kratkymu} 
deviates from eq~\ref{eq_monokratky} only by the numerical coefficient 
in front of $1/\sqrt{N}$. The difference is $\sim 10\%$.

\subsection{Infinite chain limit and scaling arguments}
\label{subsec_Scaling}

It is worthwhile to discuss the infinite chain limit, $\mu \rightarrow 0$, 
that puts forward most clearly the essential differences with an ideal chain.
(The differences between monodisperse and polydisperse systems are inessential
in this limit.) 
We can write the form factor of an infinite chain ($R_\text{g} \rightarrow
\infty$) at scales larger than $\xi$ as:
\begin{equation} \label{eq_infformfac}
F({q}) =\frac{12}{q^2 \bstar^2} \frac{1}{\left(1+\frac{3}{8}
\frac{q b^*}{\rho \hspace{0.05cm} b^{*3}}\right)} .
\end{equation}
Following  standard notations \cite{cavallowittmer,SchaferBook,LSMMKB2000},
we may rewrite eq~\ref{eq_infformfac} in the form:
\begin{equation} \label{OneLoopForm}
\frac{1}{F({q})} =\frac{q^2 \bstar^2}{12} +\frac{1}{32} \frac{q^3}{\rho}.
\end{equation}
See the Figures~\ref{fig4} (bold line), \ref{fig6} (dashed line) and \ref{fig9}
(bold line) for representations of this important limiting behavior.
The correction term obtained in the one-loop approximation (eq~\ref{OneLoopForm})
depends neither on the excluded volume parameter $v$ nor on the statistical segment. 
Hence, it is expected to hold even in the strongly fluctuating semidilute regime
and it is of interest to compare our results with the recent renormalization group 
calculations of L.~Sch\"afer \cite{SchaferBook}. There the skeleton diagrams for the 
renormalization of interaction and statistical segment have also been performed within
the one loop approximation. From the above it is expected that both results are identical. 
After careful insertions (eq~{18.23}, p.~{389} of Ref.~\cite{SchaferBook}) 
a $q^3/\rho$ correction can be extracted with the universal amplitude $0.03124..$.
The fact that this numerical amplitude is so close to our $1/32=0.03125$ 
comforts both our and Sch\"afer's result.

Performing an inverse Fourier transform of the form factor,
eq~\ref{eq_infformfac} gives not  only
the Coulomb-like term from the singularity at the origin, but also
another long-range contribution arising from the branch cut:
\begin{equation} \label{RealSpaceCorr}
G_\text{intra}({ r}) = \frac{12}{\bstar^2} \frac{1}{4 \pi r} -
\frac{9}{4} \frac{1}{\rho \bstar^4}
\frac{1}{\pi^2 r^2}.
\end{equation}
The correction is never dominant in real space. But both
contributions are different in nature. In the collective structure
factor \cite{ANSSO2005,SOANSPRL2005} 
$S^{-1}(q)\sim v+ c_2 q^2 + c_3 q^3$, the leading
singularity of eq~\ref{eq_infformfac} is shifted away from the origin
and the corresponding contribution is screened
on the lengthscale $\xi$ in real space. The  branch cut (from the
$q^3$ term) still contributes a power law, namely an anticorrelation
term decreasing as $r^{-6}$, that has been identified as a
fluctuation-induced Anti-Casimir effect \cite{ANSSO2005,SOANSPRL2005}
(or as the $n-1=-1$ Goldstone mode in the polymer-magnet analogy
\cite{SO1990}). The average number
of particles from the same molecule in a sphere of radius
$R$,  $\mathcal{N}(R)=\int G_\text{intra}({r}) d^3{\bf r}$ is decreased
(compared to a Gaussian coil) because of the sign
of the correction. Nevertheless, the differential (apparent) Hausdorff dimension
\cite{Hausdorff} $d_\text{H}$ as defined by $d_\text{H}= {\text d}\log
\mathcal{N}(R)/{\text d}\log R$ is increased.

The fluctuation corrections presented in
eqs~\ref{eq_infformfac},\ref{OneLoopForm},\ref{RealSpaceCorr}
can be interpreted with the following argument, involving $\tilde{G}(r,s)$, the
correlation function of two points separated by $s$ monomers,
%
%
$\tilde {G}(r,s) = \int G(q,s)e^{{\rm i} {\bf q} \cdot {\bf r}}
\frac{{\rm d}^3q}{(2\pi)^3}$.
For large $s$, the difference
$\delta \tilde{G}(r,s)\equiv \tilde{G}(r,s)-\tilde{G}^{(0)}(r,s)$ is discussed below 
in the limit of large and small geometrical separation $r$ between monomers. 
Here $\tilde {G}^{(0)}(r,s)={\rm c}{\rm o}{\rm n}{\rm s}{\rm
t}\,\,\exp\left(-\frac 32\frac {r^2}{s \bstar^{2}}\right)$, 
where the renormalized $\bstar$ is used instead of $b$.

A highly stretched $s$-fragment ($r \gg \bstar \sqrt{s}$) can be viewed
as a string of Pincus blobs, each blob of $g\sim s^2 b^2/r^2$ units.
Different blobs do not overlap in this limit, therefore the effective 
statistical segment of the blob, $b_\text{eff}$
comes as a result of interactions of units inside the blob 
(see eq~\ref{EffStatSeg})
\begin{equation}
b^2_{\text{eff}} = \bstar^2\left(1 - \frac{const}{\rho \hspace{0.05cm} 
\bstar^3 \sqrt{g}} \right)
\end{equation}
where $const$ is a universal numerical constant. The elastic energy 
of stretching the
$s$-segment is therefore 
\begin{equation}
\nonumber
W_{el}(r)\simeq n_b\frac 32\frac
{\left(r/n_b\right)^2}{gb_{\text{eff}}^2}=\frac 32\frac {
r^2}{sb_{\text{eff}}^2}\simeq\frac 32\frac {r^2}{s\bstar^{2}}\left(1+{\rm c}{\rm
o}{\rm n}{\rm s}
{\rm t}\,\frac r{\rho \bstar^{4}s}\right)
\nonumber
\end{equation}
where $n_b=s/g$ is the number of blobs in the $s$-segment.
Thus, when $r \gg \sqrt{s} \bstar$:
\begin{equation}
%
\tilde {G}(r,s)={\rm c}{\rm o}{\rm n}{\rm s}{\rm
t}\,e^{-W_{el}(r)}\simeq \tilde{G}^{(0)}(r,s)
\exp\left(-{\rm c}{\rm o}{\rm n}{\rm s}{\rm t}\,\frac
{r^3}{\rho \hspace{0.05cm} \bstar^{6}s^2}\right)
\end{equation}
The faster decay of $\tilde{G}$ as compared to $\tilde{G}^{(0)}$ 
leads to higher scattering
(positive $\delta F(q)$) at small $q$.

\begin{figure}[t]
\centering
\includegraphics*[width=5cm]{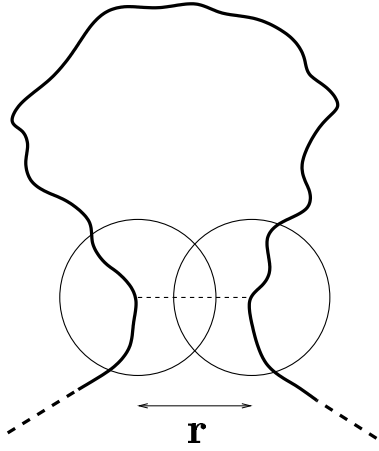}
\caption{For small internal end-to-end distances ($r \ll \sqrt{s} \bstar$),
monomers of terminal blobs with $g \sim r^2/\bstar^2$ units interact.}
\label{fig7}
\end{figure}

On smaller length scales, $r \ll \sqrt{s} \bstar$, it is convenient to consider 
the $s$-fragment as a chain of blobs of size $r$, with $g\sim r^2/{\bstar^2}$. 
This is sketched in Figure~\ref{fig7}.
The correction $\delta \tilde G$ here
is essentially due to the direct interaction of the overlapping 
blobs or radius $r$ around the two correlated monomers.
The number of binary contacts is proportional
to $g^2/r^3$, while the pairwise interaction between monomers scales 
like $1/ (g \rho)$
(see eq~\ref{potential}), giving \cite{ANSAJ2003} $U \sim 
g/(\rho\hspace{0.05cm} r^3)
\sim 1/(r \bstar^2 \rho)$.
Therefore, for $\xi \ll r \ll \sqrt{s}\hspace{0.05cm} \bstar$,
\begin{equation}
\begin{split}
\tilde{G}(r,s)&\simeq \tilde{G}^{(0)}(r,s)\left(1 - U \right) \\
               &\simeq \tilde{G}^{(0)}(r,s)\left(1-\frac{const}{r \bstar^2 \rho} \right),
\end{split}
\end{equation}
which qualitatively explains eq~\ref{RealSpaceCorr}.
For $q\gg \frac{1}{\sqrt{s} b}$, we get:
\begin{equation}\label{SmallFrag}
\delta G(q,s) = -const\left(\frac{6}{4 \pi \bstar^2 s} \right)^\frac{3}{2}
\frac{4 \pi}{q^2 \bstar^2 \rho}.
\end{equation}
This regime is also limited by the condition $q \xi \ll 1$. Thus the 
low-$q$ correction
is positive and it increases with $q$, while the high-$q$ correction
$\delta G(q,s)$ is negative and is also increasing with $q$, implying an
intermediate decline. This non-monotonic behavior of $\delta G(q,s)$
translates in a non-monotonic dependence of $\delta F(q)$ for finite $N$
(Figure~\ref{fig6}).

To underline the origin of the "non-analytical" term $1/q$, let us give some 
details of an easy calculation of the correction to the correlation function 
for the infinite chain. The top-left diagram in Figure~\ref{fig1} is the only 
diagram producing this term.  
\begin{figure}[t]
\centering
\includegraphics*[width=7cm]{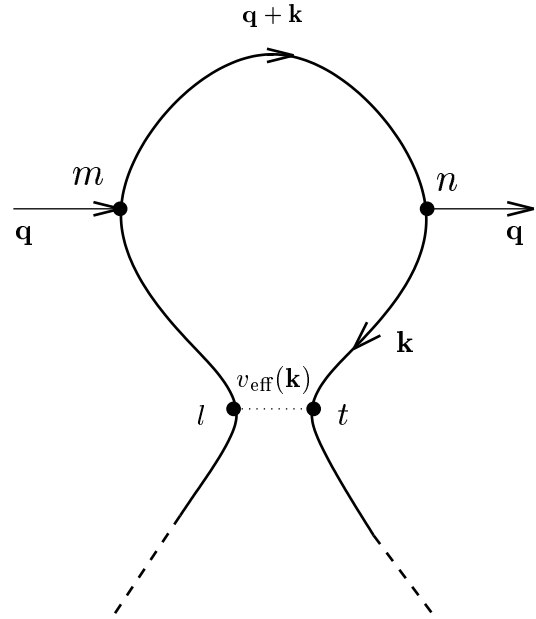}
\caption{Diagram producing the non-analytical term for an infinite chain.
The mixing of the wave-vectors is pinned in the (m,n) segment. 
The summation is done on (n,t) and (m,l).}
\label{fig8}
\end{figure}
With notations shown in Figure~\ref{fig8} 
we can write the corresponding analytical expression:
\begin{equation}
\begin{split}
\delta& F(q)=-\underset{t-n}{\sum_{n-m,m-l}}\int{d{\bf k}}\hspace{0.05cm}\veff({\bf k}) 
\exp({-{\bf k}^2(m-l)a^2})\\& 
\exp(-({\bf q}+{\bf k})^2(n-m)a^2) \exp({-{\bf k}^2(n-t)a^2}).
\end{split}
\end{equation}
Here $e^{-{\bf k}^2(m-l)a^2}$ is the Gaussian (bare) correlation function of the chain 
between monomers $l$ and $m$. In the 
interaction, we will leave only the $k^2$-dependent part. After 
summation (integration) over $m-l$ and $n-t$ we get:
\begin{equation}
\delta F(q) = -\frac{v \xi^2}{a^4} \sum_{n-m=0}^\infty \int \exp({-({\bf q}+{\bf k})^2(m-n)a^2}) d\cos(\theta) dk.
\end{equation}
Then, after integration over $n-m$ we get:
\begin{equation}
\delta F(q) = -\frac{v \xi^2}{|q|a^6}\int_{-1}^{1} \frac{d\cos(\theta)}{\sqrt{1-\cos^2(\theta)}}
\simeq -\frac{1}{\rho\bstar^3} \frac{1}{|q|\bstar} 
\end{equation} 
which is similar to eq~\ref{SmallFrag}. 
It should be emphasized that for a finite chain, the finite summation over $m-l$ 
and $n-t$ produces extra terms, and these "odd" terms exist only for $qR_\text{g}<1$.

\clearpage
\newpage
\section{Computational results}
\label{sec_comput}


The previous section contains non-trivial results due to {\em generic} physics 
which should apply to all polymer melts containing {\em long} and preferentially 
{\em flexible} chains. 
We have put these predictions to a test by means of extensive lattice 
Monte Carlo simulations \cite{BWM04} of linear polymer melts having
either a quenched and monodisperse or an annealed size-distribution. 
For the latter ``equilibrium polymers" (EP) one expects (from standard 
linear aggregation theory) a Flory distribution if the scission energy 
$E$ is constant (assuming especially chain length independence) 
\cite{CC90,WMC98,footquench2anneal}. 
(Apart from this finite scission energy for EP all our systems are perfectly 
athermal. We set $k_BT=1$ and all length scales will be given below in 
units of the lattice constant.) 
We sketch first the algorithm used and the samples obtained and discuss
then the intramolecular correlations as measured by computing the single
chain form factor $F(q)$.

\subsection{Algorithm and some technical details}
\label{sub_algo}

For both system classes we compare data obtained with the three dimensional 
bond-fluctuation model (BFM) \cite{BFM} where each monomer occupies the eight 
sites of a unit cell of a simple cubic lattice. 
For details concerning the BFM see Refs.~\cite{BFM,BWM04,WMC98}.
For all configurations we use cubic periodic simulation boxes of linear size $L=256$ 
containing $2^{20}$ monomers. These large systems are needed to suppress finite 
size effects, especially for EP \cite{WMC98}. 
The monomer number relates to a number density $\rho=1/16$ where half of the 
lattice sites are occupied (volume fraction $0.5$). It has been shown that this 
``melt density" is characterized 
by a small excluded volume correlation length $\xi$ of order of the monomer size 
\cite{BFM,BWM04}, by a low (dimensionless) compressibility, 
$S(q\rightarrow 0)/\rho=1/v\rho = 0.24$, as may be seen from the bold line ($f=1$) 
indicated in the main panel of Figure~\ref{fig10} below, and an effective statistical 
segment length, $\bstar \approx 3.24$ \cite{jojoPRL,footbstar}.
The chain monomers are connected by (at most two saturated) bonds.
Adjacent monomers are permanently linked together for monodisperse
systems (if only local moves through phase space are considered).
Equilibrium polymers on the other hand have a finite and constant 
scission energy $E$ attributed to each bond (independent of density, 
chain length and position of the bond) which has to be paid whenever
the bond between two monomers is broken. 
Standard Metropolis Monte Carlo is used to break and recombine the chains. 
Branching and formation of closed rings are explicitly forbidden and only 
linear chains are present. The delicate question of ``detailed balance", 
i.e. microscopic reversibility, of the scission-recombination step is dealt 
with in Ref.~\cite{HXCWR06}. 

The monodisperse systems have been equilibrated and sampled using a mix of local, 
slithering snake and double pivot moves extending our previous studies 
\cite{jojoPRL,papEPL}. 
This allowed us to generate ensembles containing about $10^3$ independent configurations 
with chain length up to $N=8192$ monomers. More information on equilibration and sample 
characterization will be given elsewhere \cite{papRsPs}.
We have sampled EP systems with scission energies up to $E=15$,
the largest energy corresponding to a mean chain length $\Nav \approx 6000$.
Some relevant properties obtained for these EP systems are summarized in Table~\ref{tab_EP}.
It has been verified (as shown in Refs.~\cite{WMC98,HXCWR06}) that the 
size-distribution obtained by our EP systems is indeed close to the Flory 
distribution studied in the analytical approach presented in 
the previous section. 
For EP only local hopping moves are needed in order to sample independent 
configurations since the breaking and recombination of chains reduce the 
relaxation times dramatically, at least for large scission-recombination 
attempt frequencies \cite{footequilibrate}. 
In fact, all EP systems presented here have been sampled 
within four months while the sample of monodisperse $N=8192$ configurations
alone required about three years on the same XEON processor. EP are therefore 
very interesting from the computational point of view, allowing for an 
efficient test of theoretical predictions which also apply to monodisperse systems.
%

\subsection{Form factor}
\label{sub_simuFQ}
\label{sub_simuEP}

\begin{figure}[t]
\centering
\includegraphics*[width=11cm]{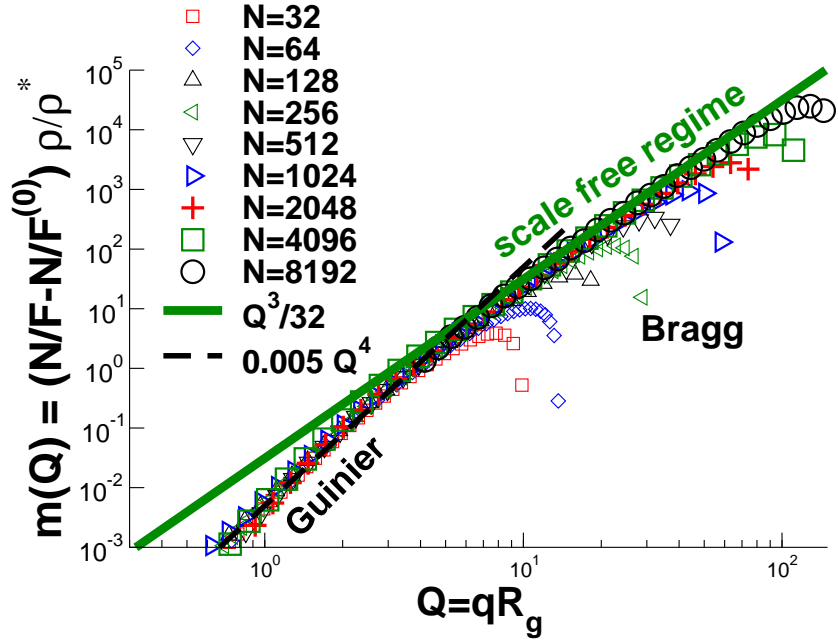}
\caption{Scaling attempt of the non-Gaussian deviations of the form factor of
monodisperse polymers in the melt in terms of the {\em measured} radius of
gyration $\Rgyr$ (instead of \Rgid). As suggested by eq~\ref{OneLoopForm},
the difference $1/F(q)-1/F^{(0)}(q)$ of the measured and the ideal
chain Debye form factor has been rescaled by the factor $N \rho/\rhostar$
and plotted as a function of $Q= q \Rgyr$.
We obtain perfect data collapse for all chain lengths included.
(Obviously, data points in the Bragg limit $q\approx 1$ do not scale.)
The difference $-\delta F(q)$ is positive in all regimes and no change
of sign occurs in this representation.
Note that the power law slope, $m(Q) = Q^3/32$, predicted by eq~\ref{OneLoopForm},
can be seen over more then one order of magnitude. In the Guinier regime,
the difference increases more rapidly, $m(Q) \propto Q^4$ (dashed line),
as one expects from a standard analytic expansion in $Q^2$.
%
%
}
\label{fig9}
\end{figure}

It is well known that the intramolecular single chain form factor of monodisperse 
polymer chains may be computed using 
\begin{equation}
F_N(q) \equiv \frac{1}{N} \la 
\left( \sum_{i=1}^N\cos({\bf q \cdot r}_{i}) \right)^2  +
\left( \sum_{i=1}^N\sin({\bf q \cdot r}_{i}) \right)^2 
\ra ,
\label{eq_defFQmono}
\end{equation}
the average being taken over all chains and configurations available.
The form factor obtained for the largest available monodisperse chain systems
currently available is represented in the 
Figures~\ref{fig4}, \ref{fig6}, \ref{fig7} and \ref{fig9}. 
It should be emphasized that the correct generalization of eq~\ref{eq_defFQmono} 
to polydisperse systems compatible with eqs~\ref{defpoly} and \ref{eq_mono2poly} 
is the average with weight $N \rho_N$ over $F_N(q)$.
In practice, one computes simply the ensemble averaged sum over 
$(\sum_{i=1}^N\sin({\bf q \cdot r}_{i}))^2 + (\sum_{i=1}^N\cos({\bf q \cdot r}_{i}))^2$ 
contributions for each chain and divides by the total number of monomers.

Figure~\ref{fig2} presents the (unscaled) form factors obtained for four 
different scission energies for our BFM EP model at $\rho=1/16$. 
The three different $q$-regimes are indicated. Details of the
size-distribution must matter most in the Guinier regime which
probes the total coil size. Non-universal contributations to the
form factor arise at large wavevectors (Bragg regime).
Obviously, the larger $E$ the wider the intermediate Kratky regime 
(see the dashed line indicating eq~\ref{eq_kratkyregime})
where chain length, polydispersity and local physics should not 
contribute much to the deviations of the form factor from ideality. 
A very similar plot (not shown) has been obtained for monodisperse polymers.
Not surprisingly, it demonstrates that the form factors of both system classes 
become indistinguishable for large wavevectors.
 
The natural scaling attempt for the form factor of EP is presented 
in Figure~\ref{fig3} for a broad range of scission energies.
We plot $F(Q)/F(0)$ as a function of $Q=q \Rgyr$
where both $F(q\rightarrow 0) = \la N^2\ra/\Nav$ and 
the (Z-averaged) gyration radius $\Rgyr$ have been measured directly for each $E$. 
Note that the strong variation of $F(0)$ and $\Rgyr$ with $E$ showing that the 
successful scaling collapse is significant. Obviously, this scaling does not hold in 
the Bragg regime ($q \approx 1$) where $F(q)$ increases rapidly, as one expects.
The bold line represents the ideal chain form factor, eq~\ref{eq_idealform}, 
where the identification of the coefficients, 
$F(0) \rightarrow 2/\mu$ and $\Rgyr^2 \rightarrow 3 a^2/\mu$,
is suggested by the Guinier limit, eq~\ref{eq_guinier}.
Hence, the perfect fit for $q \ll 1/\Rgyr$  is imposed,
but the agreement remains nice even for much larger wavevectors. 
A careful inspection of the Figure reveals, however, that eq~\ref{eq_idealform}
overestimates systematically the data in the Kratky regime.
(The corresponding plot for monodisperse chains is again very similar.)

This can be seen more clearly in the Kratky representation given in 
Figure~\ref{fig4} in linear coordinates. We present here the systems
with the longest masses currently available for both monodisperse
($N=4096$ and $N=8192$) and EP systems ($E=14$ and $E=15$). 
The  non-monotonous behavior is in striking conflict 
with Flory's hypothesis. The difference between the ideal 
Gaussian behavior (thin line) and the data becomes up to 20\%.
For the large (average) chain masses given here {\em all} 
systems are identical for $q \geq 0.1$ (but obviously not on larger scales).
It should be noted that qualitatively similar results have been reported 
--- albeit for much shorter chains --- for more than a decade in the
literature \cite{CSGK91,LSMMKB2000}.
The infinite chain prediction, eq~\ref{OneLoopForm} (or equivalently
eq~\ref{eq_infformfac}), gives a lower envelope for the data which fits
reasonably --- despite its simplicity --- in the finite wavevector range
$0.1 < q < 0.4$.

The form factor difference $\delta F(q) = F(q) - F^{(0)}(q)$ 
is further investigated in the Figures~\ref{fig5} and \ref{fig6}
for equilibrium and monodisperse systems respectively.
These plots highlight the deviations in the Guinier regime.
In both cases the ideal chain form factor $F^{(0)}(q)$ is computed
assuming the same effective statistical segment length $\bstar=\astar \sqrt{6}$,
i.e. the reference chain size is $\Rgid = \astar N^{1/2}$.
In the first case the reference $F^{(0)}(q)$ is the ideal chain form 
factor for Flory-distributed chains, eq~\ref{eq_idealform}, in the 
second the Debye function 
$ F^{(0}(q) = N f_D(Q^2)$ with $f_D(x) = (\exp(-x) - 1 + x) 2/x^2$ 
\cite{doi1989}. 

As suggested by eq~\ref{totalpert} and eq~\ref{eq_mono} respectively 
we plot $\delta F(q)/c$ {\em vs.} $Q=q \Rgid$, i.e. the axes have been 
chosen such that the data should scale for different (mean) chain length.
We obtain indeed a reasonable scaling considering that our chains are not 
large enough to suppress (for the $Q$-range represented) the deviations 
$\delta F(q)$ due to local physics. The scaling shows implicitly that
the corrections with respect to the infinite chain limit decay as
the inverse gyration radius, $1/c \sim 1/\sqrt{N}$,
as predicted by eqs~\ref{kratkymu} and \ref{eq_monokratky}.
(Both plots appear to improve systematically with increasing chain length and, 
clearly, high precision form factors for much larger chains must be considered 
in future studies to demonstrate the scaling numerically.)
Also the functional agreement with theory is qualitatively satisfactory 
in both cases, for equilibrium polymers it is even quantitative 
for small wavevectors. 
For monodisperse chains we find numerically a much more pronounced hump in 
the Guinier as the one predicted by eq~\ref{eq_mono} (bold). This is very 
likely due to the chosen interaction potential eq~\ref{potential} for 
Flory-distributed chains which is not accurate enough for the description 
of the Guinier regime of monodisperse chains.

It should be pointed out that the success of the representation of the 
non-Gaussian deviations chosen in the Figures~\ref{fig5} and \ref{fig6} 
does depend strongly on the accurate estimation of the statistical segment 
length \bstar \ of the ideal reference chains. A variation of a few 
percents breaks the scaling and leads to qualitatively different curves.
Since such a precision is normally not available (neither in simulation 
nor in experiment) it is interesting to find a more robust representation 
of the form factor deviations which does not rely on \bstar \ and allows
to detect the theoretical key predictions for long chains
(notably eq~\ref{OneLoopForm}) more readily.
Such a representation is given in Figure~\ref{fig9} for monodisperse chains. 
(A virtually indistinguishable plot has been obtained for EP.)
The reference chain size is set here by the {\em measured} radius of gyration 
$\Rgyr(N)$ (replacing the above \Rgid) which is used for rescaling the axis
and, more importantly, to compute the Debye function $F^{(0)}(q)$.
The general scaling idea is motivated by Figure~\ref{fig3},
the scaling of the vertical axis is suggested by eq~\ref{OneLoopForm}
which predicts the difference of the inverse form factors to be proportional 
to $N^0 q^3$. Without additional parameters ($\Rgyr$ is known to high precision)
we confirm the scaling of
\begin{equation}
\label{eq_Fqinvscal}
m(Q) \equiv \left( \frac{N}{F(q)} - \frac{N}{F^{(0)}(q)} \right) \rho/\rhostar 
\end{equation}
as a function of $Q=q \Rgyr$ 
with $\rhostar \equiv N/\Rgyr^3$ being the overlap density. 
Importantly, our simulations allow us to verify 
for $Q \gg 5$ the fundamentally novel $Q^3$ behavior of the master curve 
predicted by eq~\ref{OneLoopForm} and this over more than an order of magnitude!
%

\begin{figure}[t]
\centering
\includegraphics*[width=11cm]{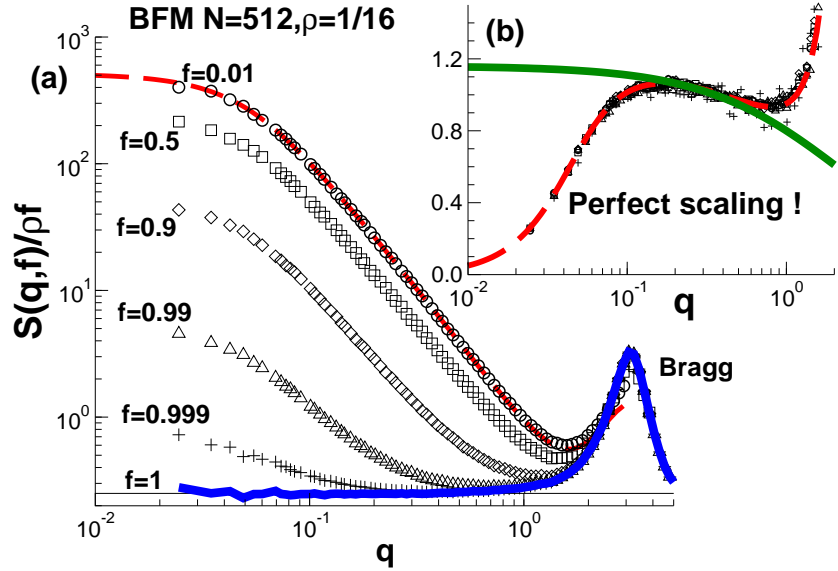}
\caption{The response function $S(q,f)$ of a fraction $f$ of marked
monodisperse chains of chain length $N=512$. The remaining $1-f$ chains
are considered to be not ``visible" for the scattering.
(The simulation box of linear size $L=256$ contains 2048 chains in total.)
The main figure {\bf (a)} presents  $S(q,f)$ directly as computed from
$S(q,f)/\rho f = \frac{1}{n} \la \left(\sum_i \cos({\bf q \cdot r}_i) \right)^2
+ \left( \sum_i \sin({\bf q \cdot r}_i) \right)^2 \ra$ where the sums run over
all $n = L^3 \rho f$ marked monomers and the wavevectors are commensurate with
the cubic box. Also included is the form factor $F(q)$
(dashed line) from eq~\ref{eq_defFQmono} which corresponds to the
$S(q,f)/\rho f \stackrel{f\rightarrow 0}{\Rightarrow} F(q) $ limit
(but compares already perfectly with the $f=0.01$ data set).
The so-called ``total structure factor" $S(q,f=1)$ (bold line at bottom)
is the Fourier transformed monomer pair-correlation function of
{\em all} monomers.
For scales larger than the monomer size it is essentially constant
and yields an accurate determination of the (dimensionless) compressibility
$S(q)/\rho = 1/v\rho \approx 0.24$. Since the compressibility is finite,
although small, this value has to be properly taken into account if
one wants to estimate the form factor $F(q)$ from the experimentally
available $S(q,f)$.
The inset {\bf (b)} presents a Kratky representation of $S(q,f)$
as suggested by an appropriate generalization of eq~\ref{labeledscatt}:
$S(q,f)-f^2S(q,1)=\rho f(1-f)F(q)$; the reduced quantity
$q^2\left[S(q,f)-f^2S(q,1)\right]/\left[\rho f(1-f)\right]$ \cite{Rubinstein,SchaferBook}
is plotted against the wavevector. 
In addition, $q^2F(q)$ obtained using eq~\ref{eq_defFQmono} is indicated by the dashed
line in the inset.
All data sets collapse perfectly which provides a striking
confirmation of eq~\ref{labeledscatt}.
Also indicated is the infinite chain asymptote eq~\ref{eq_infformfac}
(bold line).
}
\label{fig10}
\end{figure}

In this representation we do {\em not} find a change of sign for the form 
factor difference ($\delta F(q)$ is always negative) and all regimes can 
be given on the same plot in logarithmic coordinates. 
In the Guinier regime we find now $m(Q) \propto Q^4$ which is readily
explained in terms of a standard expansion in $Q^2$.
(The first two terms in $Q^0$ and $Q^2$ must vanish by construction
because of the definition of radius of gyration, 
eq~\ref{eq_guinier} \cite{doi1989}.)
Finally, we stress that the scaling of Figure~\ref{fig9} is not fundamentally
different from the one attempted in Figures~\ref{fig5} and \ref{fig6}.
Noting $F(q) F^{(0)}(q) \approx (N f_D(Q))^2$, it is equivalent to
$-\delta F(Q)/c \approx m(Q) f_D(Q)^2$ with $c \approx N^2 / \rho \Rgyr^3$.
(Compared with eqs~\ref{totalpert} and \ref{eq_mono}, 
$\Rgid$ has been replaced by the measured $\Rgyr$.)
This scaling has been verified to hold (not shown) but
we do not recommend it, 
since it does not yield simple power law regimes.

\clearpage
\newpage
\section{Conclusion}
\label{sec_conclusion}

We have shown in this paper that even for infinitely long and flexible polymer 
chains no Kratky plateau should be expected in the form factor measured from a 
dense solution or melt (see Figure~\ref{fig4}). 
We rather predict a non-monotonic correction 
to the ideal chain scattering crossing from positive in the Guinier regime 
to negative in the Kratky regime (Figures~\ref{fig5} and \ref{fig6}). 
The former regime merely depends on the radius of gyration and the correction 
corresponds to some deswelling of the coil. In the latter regime the form 
factor ultimately matches that of an infinite chain 
$\frac{1}{F({q})} =\frac{q^2 \bstar^2}{12} +\frac{1}{32} \frac{q^3}{\rho}$ 
for $q\xi \ll 1$ (Figure~\ref{fig9}).

The $q^3$-correction depends neither on the interaction nor on the
statistical segment, it must hence be generally valid, even in the
critical semidilute regime. We checked explicitly that the one-loop
correction obtained by Sch\"afer in the strongly fluctuating semidilute 
regime by numerical integration of renormalization group equations takes 
the same form with the amplitude $0.03124\ldots$ within $0.03\%$ of our $1/32$.

It is to be noted that the above correction for infinite chains
is not an analytic function of
$q^2$ as one would naively anticipate. For finite chains the correction remains 
a function of even powers of $q$. The intriguing $q^3$-correction for infinite 
chains formally arises from dilation invariance of the diagrams. Established 
theoretical methods \cite{CSGK91,Fuchs,Schulz} may implicitly {\em assume analytical} 
properties of scattering functions and non-analytical terms discussed 
in our paper could be easily overlooked.

These theoretical results are nicely confirmed by our Monte Carlo simulations
of long flexible polymers. The agreement is particulary good for
equilibrium polymers (Figure~\ref{fig5}) and satisfactory for all
systems with large (mean) chain length $N \gg 1000$ (Figure~\ref{fig4}). 
It should be emphasized that all fits presented in this paper are parameter 
free since the only model dependent parameter $\bstar$ has been independently 
obtained from the internal distances of chain segments \cite{footbstar}.
Since a sufficiently accurate value of $\bstar$ may not be available 
in general, our simulation suggest as a simple and robust way to detect 
(also experimentally) the universal $q^3$-correction the scaling 
representation of the (inverse) form factor difference in terms of 
the {\em measured} radius of gyration given in Figure~\ref{fig9}.
We expect that data for {\em any} polymer sample ---
containing long and flexible linear chains with moderate polydispersity ---
should collapse with good accuracy on the {\em same} master curve. 
Strong polydispersity 
(such as one finds in EP) should merely change its behavior in
a small regime around $Q=q \Rgyr \approx 1$. 

Measuring the form factor is a well accepted method to determine the statistical 
segment length. We already mentioned in the Introduction that there is no
clear evidence of a true Kratky plateau from experiments and further
showed that a plateau is actually not to be expected from the theory
on general grounds. We are lacking an operational definition of the
statistical segment length, even for very long flexible "thin"
chains. One way out would be in principle to fit a large $q$-range, 
from the Guinier regime --- as far it can be cleanly measured on a sample 
with controlled polydispersity --- to monomer scale, 
with the corrected formula $F^{(0)}(q)+\delta F(q)$.
However, if the size-distribution is not known precisely (as it will be 
normally the case) we recommend to determine $\bstar$ instead by means of the 
infinite chain asymptote, eq~\ref{eq_infformfac}, as can be seen Figure~\ref{fig4}.

At this point one may wonder whether eq~\ref{labeledscatt},  
(the precise form of this equation being given in the caption
of Figure~\ref{fig10})
routinely used to rationalize the scattering of a mixture of
deuterated and hydrogenated chains is accurate enough to extract the
form factor, including the corrections.  
From a theoretical point of view, for ``ideal" labeling of the chains, 
which does not introduce additional interactions between labeled and 
unlabeled chains, there is no question that this can be done. 
Practically however, there is a danger that experimental noise in 
subtracted terms in eq~\ref{labeledscatt} will mask corrections 
discussed in our paper.  
The strongest support comes here from numerical results presented 
in Figure~\ref{fig10}. We have computed the response function $S(q,f)$
for a melt of monodisperse chains for chain length $N=512$ and 
different fractions $f$ of labeled chains. 
The main panel {\bf (a)} gives $S(q,f) / f\rho$ and the form factor $F(q)$ 
as a function of the wavevector. 
The inset {\bf (b)} presents a Kratky representation of the rescaled 
structure factor: For a surprisingly large range of $f$ the data scales 
if the standard experimental procedure is followed and the scattering of 
the background density fluctuations $ f^2 S(q,f=1)$ has been properly 
substracted \cite{Rubinstein,SchaferBook}. 
The rescaled response function is identical to $F(q)$ and shows precisely the 
non-monotonic behavior, eq~\ref{eq_monokratky}, and the asymptotic infinite 
chain limit, eq~\ref{eq_infformfac} (bold line in inset), predicted by our theory.
This confirms that eq~\ref{labeledscatt} allows indeed to extract the 
correct form factor and should encourage experimentalists to revisit 
this old, but rather pivotal question of polymer science
\cite{footSQ}.

\section*{Acknowledgments}
We thank M. M\"uller (G\"ottingen) and M. Rawiso (ICS) for helpful 
discussions and J. Baschnagel for critical reading of the manuscript.
A generous grant of computer time by the IDRIS (Orsay) is also gratefully
acknowledged.

\newpage
\begin{table}[t]
\begin{tabular}{|l||c|c|c|c|c|}
\hline
$E$  & \Nav & $F(0)$ & $l$   & $\Rend$& $\Rgyr$ \\ \hline
1    & 6.4  & 11.9   & 2.632 & 12.6   & 5.2	\\
2    & 10.4 & 19.7   & 2.633 & 16.5   & 6.8     \\
3    & 16.8 & 32.4   & 2.633 & 21.6   & 8.8	\\
4    & 27.5 & 53.4   & 2.634 & 28.1   & 11.4    \\
5    & 44.9 & 87.9   & 2.634 & 36.3   & 14.8	\\
6    & 73.7 & 145    & 2.634 & 46.9   & 19.1    \\
7    & 121  & 239    & 2.634 & 60.7   & 24.7	\\
8    & 199  & 394    & 2.634 & 77.9   & 31.8    \\
9    & 328  & 650    & 2.634 & 102    & 41.4	\\
10   & 538  & 1075   & 2.634 & 129    & 52.7	\\
11   & 887  & 1766   & 2.634 & 165    & 67.7	\\
12   & 1453 & 4747   & 2.634 & 217    & 88.1    \\
13   & 2390 & 4747   & 2.634 & 270    & 110	\\
14   & 3911 & 7868   & 2.634 & 348    & 143	\\
15   & 6183 & 12272  & 2.634 & 426    & 184	\\
\hline
\end{tabular}
\vspace*{0.5cm}
\caption[]{Various properties of linear equilibrium polymers obtained 
with the BFM algorithm at number density $\rho = 0.5/8$ (volume fraction $0.5$):
imposed scission energy $E$,
the mean chain length $\Nav$, 
the ratio $F(0) = \la N^2 \ra/\Nav$ comparing the first and the
second moment of the number distribution,
the mean-squared bond length $l^2 = \la {\bf l}^2 \ra$, 
the z-averaged end-to-end distance $\Rend$ and 
the radius of gyration $\Rgyr$.
For all scission energies we have used periodic simulation boxes of 
linear size $L=256$ containing $n_{mon} = 2^{20}$ monomers.
Averages are performed over all chains and $1000$ configurations. 
\label{tab_EP}}
\end{table}

\clearpage
\newpage 
\input{bib}


\clearpage
\newpage
\section*{\large For Table of Contents Use Only}
\vspace*{2.0cm}

\hspace*{-0.5cm}\parbox{8.0cm}{
{\bf Intramolecular Form Factor in Dense Polymer Systems:
Systematic Deviations from the Debye formula}
\\

P. Beckrich, A. Johner, A. N. Semenov, S. P. Obukhov, 
H. Beno\^\i t and J. P. Wittmer

}
\parbox{0.2cm}{\hspace*{0.1cm}}
\parbox{8.9cm}{
\centerline{\resizebox{8.9cm}{!}{\includegraphics*{figTOC.eps}}}
}

\end{document}

%% file: bib.tex
%
%

%% file: draft.bbl
\begin{thebibliography}{99}

\bibitem{Flory}
Flory, P.J., \textit{Statistical Mechanics of Chain Molecules}
(Oxford University Press, New York, 1988).

\bibitem{scalingPGG}
De Gennes, P.-G., \textit{Scaling Concepts in Polymer Physics}
(Cornell University, Ithaca, N.Y., 1979).

\bibitem{doi1989}
Doi, M.; Edwards, S.F., \textit{The Theory of Polymer Dynamics}
(Clarendon Press, Oxford, 1986).

\bibitem{khokhlov+grosberg}
Grosberg, A.Y.; Khokhlov, A.R.,
\textit{Statistical Physics of Macromolecules}
(AIP Press, New York, 1994).

\bibitem{Rubinstein}
Rubinstein, M.; Colby, R.H., \textit{Polymer Physics}
(Oxford University Press, Oxford, 2003).

\bibitem{BenBook}
Higgins, J.S.; Beno\^{i}t, H.C.,
\textit{Polymers and Neutron Scattering}
(Oxford University Press, 1996).

\bibitem{RawisoLectures}
Rawiso, M.,
\textit{Journal de Physique IV} {\bf 1999}, 9, 147.

\bibitem{Boue}
Bou\'e, F.; Nierlich, M.; Leibler, L.,
\textit{Polymer} {\bf 1982}, 23, 29.

\bibitem{footInhomogeneities}
Unwanted inhomogeneities (dusts or bubbles) scatter at low-$q$,
also polydispersity effects are most important there.

\bibitem{MRRDCP1987}
Rawiso, M.; Duplessix, R.; Picot, C.,
\textit{Macromolecules} {\bf 1987}, 20, 630.

\bibitem{footqrange}
Obviously, these operational problems may be overcome in the future by using
very long {\em and} flexible polymers provided labelled and unlabelled chains
do not demix, $\chi N <2$ with $\chi$ the Flory parameter. For a blend
of hydogenated and fully deuterated polystyrene at $453 \text{K}$,
$\chi = 1.5 \cdot 10^{-4}$ \cite{Bates}.

\bibitem{Bates}
Bates, F.S.; Wignall, G.D.,
\textit{Physical Review Letter} {\bf 1986}, 57, 1429.

\bibitem{ANSAJ2003}
Semenov, A.N.; Johner, A., \textit{Eur. Phys. J. E} {\bf 2003} 12, 469.

\bibitem{jojoPRL}
Wittmer, J.P.; Meyer, H.; Baschnagel, J.; Johner, A.; Obukhov, S.P.;
Mattioni, L.; M\"uller, M.; Semenov, A.N.;
\textit{Phys. Rev. Lett.} {\bf 2004}, 93, 147801.

\bibitem{ANSSO2005}
Semenov, A.N.; Obukhov, S.P.,
\textit{J. Phys.: Condens. Matter} {\bf 2005}, 17, S1747.

\bibitem{SOANSPRL2005}
Obukhov, S.P.; Semenov, A.N.,
\textit{Phys. Rev. Lett.} {\bf 2005}, 95, 038305.

\bibitem{papEPL}
Wittmer, J.P.; Beckrich, P.; Johner, A.; Semenov, A.N.; 
Obukhov, S.P.; Meyer, H.; Baschnagel, J., 
\textit{Europhysics Letters} {\bf 2007}, accepted.

\bibitem{BeckrichThesis}
Beckrich, P.; {\em Correlation properties of linear polymers in the bulk
and near interfaces}, PhD thesis, Universit\'e Louis Pasteur, Strasbourg, France {\bf 2006}.

\bibitem{cavallowittmer}
Cavallo, A.; M\"uller, M.; Wittmer, J.P.; Johner, A.,
\textit{J. Phys.: Condens. Matter} {\bf 2005}, 17, 1697.

\bibitem{CSGK91}
Curro, J.G.; Schweizer, K.S.; Grest, G.S; Kremer, K.,
\textit{J. Chem. Phys.} {\bf 1991}, 91, 1359.

\bibitem{Auhl03}
Auhl, R.; Everaers, R.; Grest, G.S.; Kremer, K.; Plimpton, S.J,
\textit{J. Chem. Phys.} {\bf 2003}, 119, 12718.

\bibitem{Sommer05}
Sommer, J.-U.; Saalwachter, K.,
\textit{EPJ E} {\bf 2005}, 18, 167.

\bibitem{SGE05}
Svaneborg, C.; Grest, G.S.; Everaers, R.,
\textit{Europhys. Lett.} {\bf 2005}, 72, 760.

\bibitem{footMFcycles}
A more subtle effect arises from the mean-field treatment
(implicitly allowing for cycles \cite{ANSSO2005,SOANSPRL2005}) of the bath
surrounding the chain under consideration. This can be shown to be negligible.

\bibitem{BFM}
Carmesin, I.; Kremer, K.,
\textit{Macromolecules} {\bf 1988}, 21, 2819;
Paul, W.; Binder, K.; Heermann, D.; Kremer, K., 
\textit{J. Phys. II} {1991}, 1, 37.

\bibitem{BWM04}
Baschnagel, J.; Wittmer, J.P.; Meyer, H.,
\textit{Monte Carlo Simulation of Polymers: Coarse-Grained Models},
in \textit{Computational Soft Matter: From Synthetic Polymers to Proteins}
edited by N. Attig et al. (NIC Series, Volume 23, J\"ulich, 2004), pp.~83-140.

\bibitem{CC90}
Cates, M.E.; Candau, S.J.; 
{\em J.~Phys. Cond. Matt} {\bf 1990}, 2, 6869.

\bibitem{WMC98}
Wittmer, J.P.; Milchev, A.; Cates, M.E.,
{\em J.Chem. Phys.} {\bf 1998}, 109, 834.

\bibitem{HXCWR06}
Huang, C.C.; Xu, H.; Crevel, F.; Wittmer, J.P.; Ryckaert, J.-P.,
Lect. Notes Phys. (Springer) {\bf 2006}, 704, 379;  cond-mat/0604279.

\bibitem{footquench2anneal}
Note that there is no difference between the annealed and the
corresponding quenched polydispersity for infinite macroscopically
homogeneous systems as long as equilibrium properties (static rather
than dynamic properties) are concerned.  This follows from the
well-known behavior of fluctuations of extensive parameters (like
mean molecular weight, or polydispersity degree) in macroscopic
systems:  the relative fluctuations vanish as $1/\sqrt {V}$ as the total
volume $V\to\infty$.

\bibitem{RPAEd1}
Edwards, S.F.,
\textit{Proc. Phys. Soc.} {\bf 1965}, 85, 613.

\bibitem{RPAEd2}
Edwards, S.F.,
\textit{Proc. Phys. Soc.} {\bf 1966}, 88, 265.

\bibitem{cloiz}
Des Cloizeaux, J.; Jannink, G.,
\textit{Polymers in Solution : their Modelling and Structure}
(Clarendon Press, Oxford, 1990).

\bibitem{Nozieres}
Pines, D.; Nozieres, P.,
\textit{The Theory of Quantum Liquids Vol.I}
(W.A. Benjamin, Inc, New York, 1966).

\bibitem{ON83}
Ohta, T.; Nakanishi, A., \textit{J. Phys. A: Math. Gen.} {\bf 1983}, 16, 4155.

\bibitem{Duplantier86}
Duplantier, B.,
\textit{J. Stat. Phys.} {\bf 1986}, 47, 1633.

\bibitem{E75}
Edwards, S.F.,  J.Phys.A: Math.Gen. {\bf 1975}, 8, 1670.

\bibitem{ME82}
Muthukumar, M.; Edwards, S.E., \textit{J. Chem. Phys.} {\bf 1982}, 76, 2720.


\bibitem{NO81}
Nikomarov, E.S.; Obukhov, S.P.,
\textit{Sov. Phys. --- JETP} {\bf 1981}, 53, 328.

\bibitem{footcumbersome}
The full cumbersome expression leading to eq~\ref{totalpert} is not given here.

\bibitem{LiqCrysSem}
Khoklov, A.R.; Semenov, A.N.,
\textit{Journal of Statistical Physics}
{\bf 1985}, 38, 161.

\bibitem{abramowitz}
Abramowitz, M.; Stegun, I.A.,
\textit{Handbook of Mathematical Functions}
(Dover, New York, 1964).

\bibitem{SchaferBook}
Sch\"afer, L.,
\textit{Excluded Volume Effects in Polymer Solutions}
(Springer-Verlag, New York, 1999).

\bibitem{LSMMKB2000}
Sch\"afer, L.; M\"uller, M.; Binder, K.,
\textit{Macromolecules} {\bf 2000}, 33, 4568.

\bibitem{SO1990}
Obukhov, S.P.,
\textit{Phys. Rev. Lett.} {\bf 1990}, 65, 1395.

\bibitem{Hausdorff}
Berge, P.; Pomeau, Y.; Vidal, C.,
\textit{Order within Chaos}
(Hermann, Paris, 1988).



\bibitem{footPade}
This result has been cross-checked by means of a direct perturbation
calculation for monodisperse chains using the Pad\'e approximation of
Debye's formula for the effective interaction potential.


\bibitem{footbstar}
As indicated in \cite{jojoPRL} \bstar \ may be best obtained from the 
intramolecular (mean-squared) distance $\la R^2(s) \ra$ averaged
over all monomer pairs $(n,m=n+s)$ of the chains. As suggested by eq~\ref{EffStatSeg}
one plots $y= \la R^2(s) \ra/s$ as a function of $x =1/\sqrt{s}$ 
which allows the simple one-parameter fit:
$$y = \bstar \left( 1 + \frac{\sqrt{24/\pi^3}}{\rho \bstar^3} x \right).$$
The prefactor --- derived in Ref.~\cite{jojoPRL}, eq~2., for monodisperse
chains --- does also apply to polydisperse systems provided $\Nav \gg s$.
%
Note that it is in principle also possible to obtain \bstar \
from the total coil size as indicated by eqs~\ref{gyration} and
\ref{endtoend} for the polydisperse case. Due to chain end effects
it turns out that this requires much larger (mean) chain lengths 
than the recommended method above.


\bibitem{papRsPs}
Wittmer, J.P., {\em et al.}, in preparation.

\bibitem{footequilibrate}
Since the EP chains break and recombine permanently the relaxation time
of the system is not set by the typical EP radius of gyration but rather 
by the size of a small chain segment which has just about the time to diffuse 
over its radius before it breaks or recombines \cite{CC90}. The high frequency 
for the scission-recombination attempts used in our simulations ensures that
the effective recombination time is small and the dynamics is, hence,
always of Rouse-type. 
%
It should be emphasized that due to the permanent recombination events a
data structure based on a topologically ordered intra chain interactions is
not appropriate and straight-forward pointer lists between connected monomers
are required \cite{WMC98}. The attempt frequency should not by taken too
large to avoid useless immediate recombination of the same monomers and
some time must be given for the monomers to diffuse over a couple of
monomer diameters between scission-recombination attempts \cite{HXCWR06}.

\bibitem{Fuchs}
Fuchs, M.,
\textit{Z. Phys. B} {\bf 1997}, 521-530.

\bibitem{Schulz}
Schulz, M.; Frisch, H.L.; Reineker, P.,
\textit{New Journal of Physics} {\bf 2004}, 6, 77.

\bibitem{footSQ}
We have computed the response function $S(q,f=1)$ for monodisperse polymers
over a large range of densities and for a nouvel version of the BFM with 
finite overlap energies. This has been done to verify the recent prediction 
of fluctuation induced long-range repulsions between solid objects in polymer 
media \cite{ANSSO2005,SOANSPRL2005}.  
This approach suggests a systematic violation of the RPA eq~\ref{eq_RPA}
proportional to $q^3$ which we have put to a numerical test. Our findings --- 
complicated by the fact that trivial monomer-monomer correlations of 
Percus-Yevick type \cite{Fuchs,Schulz} must be correctly taken into account \cite{footMFcycles} --- 
will be presented elsewhere. 
Please note that these corrections to eq~\ref{eq_RPA} 
do correspond to higher order deviations which can be shown
to be negligible for the questions addressed in this paper.



\end{thebibliography}
